\documentclass[11pt]{article}
\usepackage{cite}
\usepackage{amsmath,amsfonts,amssymb}
\usepackage[small,bf,hang]{caption}
\usepackage{slashed}

\def\hybrid{
        \topmargin -20pt
        \oddsidemargin 0pt
        \headheight 0pt \headsep 0pt
        \textwidth 6.25in 
        \textheight 9.5in 
        \marginparwidth .875in
        \parskip 5pt plus 1pt \jot = 1.5ex}

\hybrid

 \csname
@addtoreset\endcsname{equation}{section}

\newcommand{\p}{\partial}

\def\moth{\mathsurround=0pt}
\newdimen\zo \zo=0pt

\def\tick{\leaders\hrule height 0.5ex depth 0pt \hskip 0.5pt}
\def\upboxfill{$\moth \setbox\zo\hbox{\tick}%
  \hskip 3pt\hbox to 0pt{$\tick$\hss}\hrulefill \hbox to 7.5pt{$\tick$\hss}$}

\def\dtick{\leaders\hrule height .34pt depth 0.5ex \hskip 0.5pt}
\def\downboxfill{$\moth \setbox\zo\hbox{\dtick}%
  \hskip 2pt\hbox to 0pt{$\dtick$\hss}\hrulefill \hbox to 2pt{$\dtick$\hss}$}

\def\f#1#2{{\textstyle{#1\over#2}}}

\def\bec{\begin{center}}
\def\ec{\end{center}}

\def\x{\xi}

\def\be{\begin{equation}}
\def\ee{\end{equation}}
\def\bea{\begin{eqnarray}}
\def\eea{\end{eqnarray}}
\def\ba{\begin{array}}
\def\ea{\end{array}}

\begin{document}

\begin{titlepage}
\rightline{July 2014}
\rightline{\tt MIT-CTP-4562}    
\begin{center}
\vskip 2.5cm

{\Large \bf {Green-Schwarz mechanism and $\alpha'$-deformed Courant brackets}}\\

 \vskip 2.0cm
{\large {Olaf Hohm and Barton Zwiebach}}
\vskip 0.5cm
{\it {${}^1$Center for Theoretical Physics}}\\
{\it {Massachusetts Institute of Technology}}\\
{\it {Cambridge, MA 02139, USA}}\\
ohohm@mit.edu, zwiebach@mit.edu

\vskip 2.5cm
{\bf Abstract}

\end{center}

\vskip 0.5cm

\noindent
\begin{narrower}

\baselineskip15pt

We establish that the unusual two-form gauge transformations 
needed in the Green-Schwarz anomaly cancellation
mechanism fit naturally into  an $\alpha'$-deformed
generalized geometry.   The algebra of gauge transformations
is a consistent deformation of the Courant bracket 
and features a nontrivial modification of the diffeomorphism
group.  This extension of generalized geometry 
emerged from a 
 `doubled $\alpha'$-geometry', which provides a construction 
of exactly gauge  and T-duality invariant $\alpha'$ corrections 
to the effective action.

\end{narrower}

\end{titlepage}

\newpage

\setcounter{tocdepth}{1}
\tableofcontents
\baselineskip=15pt

\section{Introduction}\setcounter{equation}{0}
In this paper we study a deformation of the Courant bracket 
of generalized geometry that emerged in $\alpha'$ 
deformations of double field theory  \cite{Hohm:2013jaa},   
and relate it to the Green-Schwarz mechanism of anomaly 
cancellation~\cite{Green:1984sg}. 
The construction of \cite{Hohm:2013jaa} extends   
the original two-derivative effective field theory 
by higher-derivative corrections that 
describe 
the {\em classical} stringy geometry of the space-time theory.  Indeed, while
the Green-Schwarz mechanism uses a novel transformation of the
antisymmetric tensor field to cancel a quantum anomaly of the space-time theory,
this transformation is needed to cancel a one-loop world-sheet
anomaly at genus zero~\cite{Hull:1985jv,Sen:1985tq}. 
Therefore the 
modified gauge transformation
is a feature of the classical space-time theory.  

 The `doubled $\alpha'$ geometry' of \cite{Hohm:2013jaa} provides an exact deformation 
of the gauge structure of double field theory (DFT) \cite{Siegel:1993th,Hull:2009mi,Hull:2009zb,Hohm:2010jy,Hohm:2010pp,Hohm:2013bwa}  
by terms of ${\cal O}(\alpha')$ in the gauge algebra and up to  
${\cal O}(\alpha'^2)$ in the gauge transformations and the invariant action.
The action contains up to six derivatives in terms of a novel `double metric' field and 
is exactly gauge invariant under the deformed gauge transformations. 
The 
relation to conventional 
actions written in terms of the metric $g_{ij}$ and the Kalb-Ramond field $b_{ij}$ has not yet been established beyond 
two derivatives. It was conjectured in \cite{Hohm:2013jaa} 
that this theory 
encodes part of the $\alpha'$ corrections of 
bosonic string theory but we  explain here  
that  it actually   
 encodes a subsector of heterotic string theory.  
Specifically, this deformation encodes the gauge 
transformations implied by Green-Schwarz anomaly cancellation in heterotic string theory that modifies 
the three-form curvature of the $b$-field by a gravitational Chern-Simons term \cite{Green:1984sg,Hull:1985jv,Sen:1985tq}. 
We show here that 
this leads to a gauge algebra 
that corresponds to a deformation of the Courant bracket
of generalized geometry \cite{Tcourant}.

The two-derivative DFT is defined on a doubled space and governed by the `C-bracket' that in turn
is a T-duality covariant extension of the Courant bracket of 
generalized geometry~\cite{Siegel:1993th,Hull:2009zb}. 
In DFT a generalized vector $V^M$, with $O(D,D)$ indices $M,N=1,\ldots,2D$, decomposes    
as $V^M=(\tilde{V}_i,V^i)$, with a vector $V$ and a one-form $\tilde{V}$, when restricted to the `physical' 
$D$-dimensional subspace of the doubled space. In this case, the pair of vector and one-form 
can be viewed as a section $V+\tilde{V}$ in $T\oplus T^*$, the direct sum of the
tangent and co-tangent bundles.   
On any $D$-dimensional 
physical subspace, the C-bracket reduces to the Courant bracket. 
The $\alpha'$ deformation of  the C-bracket  yields a non-trivial deformation of the Courant bracket on the physical subspace~\cite{Hohm:2013jaa}.
We show that an exact realization of this bracket is given by the deformed gauge transformations of
the $b$-field according to the Green-Schwarz mechanism. Conventionally, the gauge transformations 
of the Green-Schwarz mechanism
are presented as deformed local Lorentz transformations, 
but we can also realize them as 
deformed diffeomorphisms. As a central 
result of this note we 
give the deformed diffeomorphisms on the two-form $b = \f{1}{2} b_{ij} dx^i\wedge dx^j$ and show that they close according to 
the deformed Courant bracket. 
The gauge transformations of the metric are unchanged and the gauge transformations 
of $b$ read 
 \be\label{nonlinDiff2Intro}
   \delta_{\xi+\tilde{\xi}}\, b
   \ = \ {\rm d}\tilde{\xi}\, +\, {\cal L}_{\xi}b\, +\, \f{1}{2} \, \hbox{tr} \bigl(  {\rm d} (\partial \xi) \wedge \Gamma \bigr)\;, 
 \ee
 with $\tilde{\xi}$ the one-form parameter, ${\cal L}_{\xi}$  the Lie derivative 
along the vector $\xi$, and $\Gamma$ the Christoffel one-form connection.   The component
version of this equation is given in (\ref{nonlinDiff2}). 
The gauge algebra of these deformed transformations is governed by 
the deformed Courant bracket $ [ \cdot \, , \, \cdot ]' $ defined by  
  \be\label{CourantIntro}
 \begin{split}
\phantom{\Bigl(}   \Big[\, \xi_1+\tilde{\xi}_1, \xi_2+\tilde{\xi}_2 \, \Big]'  \ = \ \  &
  \big[\,\xi_1\,,\xi_2\, \big]\, +\, {\cal L}_{\xi_1}\tilde{\xi}_2-{\cal L}_{\xi_2}\tilde{\xi}_1
  -\f{1}{2} {\rm d}\big(i_{\xi_1}\tilde{\xi}_2-i_{\xi_2}\tilde{\xi}_1\big) \ \  \\
  &\hskip-8pt -\f{1}{2}\big(\tilde\varphi(\xi_1,\xi_2)-\tilde\varphi(\xi_2,\xi_1)\big) \;, 
 \end{split}  
 \ee 
where we defined the map $\tilde{\varphi}$ that, given any two vectors $V$ and $W$, produces a `one-form', 
\be\label{phtildeIntro}
\tilde \varphi (V,W) \ \equiv    \
{\rm tr}\big({\rm d} (\partial V)\partial W\big)  \equiv \ {\rm tr}\big(\partial_i \partial V\partial W\big) dx^i \ \equiv \ 
\partial_i\partial_kV^l \partial_lW^k dx^i\;. 
\ee
The first line in (\ref{CourantIntro}) 
defines the standard 
Courant bracket.  
The first term on the right-hand side is the 
Lie bracket of two vector fields and defines the vector part of the bracket, while the remaining terms define the 
one-form part of the bracket. Here $i_{\xi_1}\tilde{\xi}_2$  
denotes the natural pairing between vectors and one-forms. 
The second line in (\ref{CourantIntro}) 
is the deformation of the bracket. 
Note that $\tilde{\varphi}$ is not a genuine  
one-form as the partial derivatives of vectors are not 
tensors;  
$\tilde{\varphi}$ has an anomalous transformation 
under diffeomorphisms.  
While this deformation of the Courant bracket 
is not  
diffeomorphism covariant, there is a deformed 
notion of diffeomorphisms, with respect to which the 
deformed Courant bracket \textit{is} covariant.  
We note that the exact term on the first line of (\ref{CourantIntro})
is 
not determined by closure of the gauge transformations on $g$ and $b$.
It is fixed, instead,  by the requirement that transformations called $B$-shifts 
are automorphisms of the bracket \cite{Severa:2001qm,Gualtieri:2003dx}. 
These transformations  
change $b$ by the addition of a closed two-form $B$ and act on the one-form gauge parameter as well. 
The $B$-shifts are also automorphisms of the  $[\;,\;]'$  bracket.  Let us finally note that 
structures closely related to the construction of \cite{Hohm:2013jaa} have been discussed in 
\cite{Nekrasov,Schulgin:2013xya}. Ref.~\cite{Nekrasov} obtained Courant structures in the context of 
a $\beta\gamma$-system and computed world-sheet loop corrections as in \cite{Hohm:2013jaa}.  Worldsheet vertex operator algebras that exhibit $\alpha'$ corrections
have also been investigated in \cite{Schulgin:2013xya}.

This note is organized as follows. In sec.~2 we analyze the gauge transformations of \cite{Hohm:2013jaa} 
applied to  the metric and $b$-field fluctuations to linearized order about a background. 
We show that, up to field and parameter 
redefinitions, they agree with those of the Green-Schwarz mechanism to linear order. 
The Green-Schwarz mechanism  involves 
deformed local Lorentz 
transformations and Lorentz-Chern-Simons modifications of the field strength. 
In order to relate them to the deformed Courant bracket, we recast this formulation 
in terms of deformed diffeomorphisms and Chern-Simons modifications based on 
Christoffel symbols. These transformations close exactly according to the 
 deformed Courant bracket. 
Next, in sec.~3, we review the relation between the 
undeformed C-bracket and Courant bracket. In particular, we use the opportunity to discuss 
how $B$-shifts are realized 
on the C-bracket.   
Finally, in sec.~4 we discuss the $\alpha'$-corrected C-bracket of \cite{Hohm:2013jaa}
from which the deformed Courant bracket $[\;,\;]'$ arises. We give the deformed diffeomorphisms on the one-form
and prove in a self-contained
fashion the covariance of the deformed bracket under deformed diffeomorphisms introducing some useful notation.

We close with an outlook in sec.~5. In particular we discuss more general $\alpha'$ corrections, 
relevant both for bosonic and heterotic string theory. 
This will be considered in detail in an 
upcoming paper \cite{HZupcoming}. 
In appendix A we discuss the issues associated with the finite form of the deformed 
diffeomorphisms, and in appendix B we present some details of the proof of covariance 
of the deformed Courant bracket.

\section{Green-Schwarz mechanism and deformed diffeomorphisms}

In this section we analyze perturbatively the gauge transformations of the `doubled $\alpha'$ geometry'
in terms of the fluctuations of the metric and $b$-field. We perform the field redefinitions needed to 
show that the metric fluctuation transforms conventionally but that the $b$-field fluctuation 
receives a non-trivial modification in agreement with the Green-Schwarz mechanism. 
Then we give a non-linear extension as deformed diffeomorphism transformations 
on the $b$-field and show that they close according to a deformed bracket.

\subsection{Perturbative clues}  \label{pert-clues} 
We start from the gauge transformations derived in \cite{Hohm:2013jaa} specialized to the fluctuations 
around a constant background, to the order relevant for a cubic action. The detailed derivation 
of these transformations will be presented in \cite{HZupcoming}. 
Projecting to the symmetric part $h_{ij}$ and the antisymmetric part $b_{ij}$ of the fluctuation, 
respectively, one finds deformed gauge transformations of the form    
 \be\label{defdelta1}
  \begin{split}
   \delta^{} h_{ij} \ &= \ \partial_i\epsilon_j+\partial_j\epsilon_i+  \f{1}{2} \big( \partial^kh^{l}{}_{j}\,\partial_i\partial_{[k}\,\tilde{\epsilon}_{l]}
   +\partial^kb_{i}{}^{l}\,\partial_j\partial_{[k}\,\epsilon_{l]}+(i\leftrightarrow j)\big) \;, \\
    \delta^{} b_{ij} \ &= \ \partial_i\tilde{\epsilon}_j-\partial_j\tilde{\epsilon}_i
    - \f{1}{2}\big(\partial^kh^{l}{}_{j}\,\partial_i\partial_{[k}\,\epsilon_{l]}
   -\partial^kb^l{}_{j}\,\partial_i\partial_{[k}\,\tilde{\epsilon}_{l]}-(i\leftrightarrow j)\big) \;. 
  \end{split}
 \ee  
Here $\epsilon_i$ and $\tilde{\epsilon}_i$ are the diffeomorphism and $b$-field gauge parameter, respectively, 
for the linearized gauge transformations. 
In this section 
we will consistently omit terms that are of zeroth order in $\alpha'$ and linear in fields, 
as these are irrelevant for our analysis. We also set $\alpha'=1$ as the ${\cal O}(\alpha')$ corrections 
are readily recognized by their higher derivatives. 
In (\ref{defdelta1}) we have a higher-derivative deformation that is not present for standard Einstein variables.  
We now 
ask to what extent these deformations of the gauge transformations can be 
removed by a field and/or parameter redefinition. For the symmetric part of the fluctuation this is 
indeed possible by redefining 
  \be
   h_{ij}' \ = \ h_{ij} + \f{1}{2}\big( \partial^kh^{l}{}_{i}\,\partial_{[k}b_{l]j}+(i\leftrightarrow j)\big) \;. 
  \ee
To this order this leads to extra transformations  $\delta^{1}$ from the lowest-order, inhomogeneous  
variations in (\ref{defdelta1}) of the higher-derivative terms. We compute  
   \be
 \begin{split}
 \delta^1h_{ij}' \ &= \ \delta h_{ij} +\f{1}{2}\big(\partial^k(\partial^{l}\epsilon_{i}+\partial_i\epsilon^l)\,\partial_{[k}b_{l]j}
 +\partial^{[k}h^{l]}{}_{i}\,\partial_{k}(\partial_l\tilde{\epsilon}_j-\partial_j\tilde{\epsilon}_l)+(i\leftrightarrow j)\big) \\
 \ &= \  \delta h_{ij} - \f{1}{2}\big(\partial^kh^{l}{}_{j}\,\partial_i\partial_{[k}\,\tilde{\epsilon}_{l]}
   +\partial^kb_{i}{}^{l}\,\partial_j\partial_{[k}\,\epsilon_{l]}+(i\leftrightarrow j)\big)\;. 
 \end{split}
 \ee  
Comparing with the first equation in (\ref{defdelta1}) we infer that the higher-derivative terms are 
precisely cancelled. This proves that the deformed gauge transformation for the symmetric part of the 
fluctuation is trivial and thus removable by a field redefinition. 
Let us now turn to the antisymmetric part of the fluctuation. The second term for $\delta b$ 
in (\ref{defdelta1}) can be removed by 
a combined field and parameter redefinition. In general we may perform a field-dependent 
redefinition of $\tilde{\epsilon}_i$, 
  \be\label{PARAMRedfi}
  \tilde{\epsilon}_i' \ = \ \tilde{\epsilon}_i+\Delta_i(h,b,\epsilon) \;, 
 \ee
to arrive at an equivalent  
 modified gauge transformation 
 \be
  \tilde{\delta} b_{ij} \ \equiv  \ \partial_i\Delta_j-\partial_j\Delta_i+\delta b_{ij} \;.  
 \ee  
This shows that we can simply `integrate by parts' $\partial_i$ and $\partial_j$ derivatives in $\delta b_{ij}$ exploiting possible parameter redefinitions. 
Therefore, $\delta b_{ij}$ in (\ref{defdelta1}) is equivalent to 
 \be
    \delta^{} b_{ij} \ = \ \partial_i\tilde{\epsilon}_j-\partial_j\tilde{\epsilon}_i
    +\partial^kh^{l}{}_{[i}\,\partial_{j]}\partial_{[k}\,\epsilon_{l]}
   -\partial_{[i}\partial^kb^l{}_{j]}\,\partial_{[k}\,\tilde{\epsilon}_{l]}\;. 
 \ee
Due to the antisymmetry in $i,j$ and $k,l$, 
the last term can be rewritten in terms of the three-form curvature $H_{ijk}=3\partial_{[i}b_{jk]}$, 
 \be\label{reDEFB}
    \delta^{} b_{ij} \ = \ \partial_i\tilde{\epsilon}_j-\partial_j\tilde{\epsilon}_i
    +\partial^kh^{l}{}_{[i}\,\partial_{j]}\partial_{[k}\,\epsilon_{l]}
   -\f{1}{2}\,\partial_{[i}H_{j]kl}\,\partial^{[k}\,\tilde{\epsilon}^{l]}\;. 
 \ee
Performing next the field redefinition 
 \be
  b_{ij}' \ = \ b_{ij}+\f{1}{4}\,\partial_{[i}H_{j]kl}\,b^{kl}\;, 
 \ee 
it is manifest from the gauge invariance of $H$ that the induced extra variation 
precisely cancels the final term in (\ref{reDEFB}).   
Dropping the prime from now on, we have obtained: 
  \be\label{finaldel1}
     \delta b_{ij} \ = \ \partial_i\tilde{\epsilon}_j-\partial_j\tilde{\epsilon}_i
     +\partial^kh^{l}{}_{[i}\,\partial_{j]}\partial_{[k}\,\epsilon_{l]}
       \;. 
  \ee   
Introducing the linearized spin connection 
 \be
  \omega_{j,kl}^{(1)}  \ \equiv \ -\,\partial_{[k}\, h_{l]j}\;,   
 \ee
where at the linearized level the background vielbein $e_i{}^a=\delta_i{}^a$ allows one to identify 
curved and flat indices, this can be written as
 \be\label{finaldel2}
     \delta b_{ij} \ = \ \partial_i\tilde{\epsilon}_j-\partial_j\tilde{\epsilon}_i
     +\partial^{}_{[i}\partial^{k}\epsilon^{l}\,\omega_{j]kl}^{(1)}   \;. 
  \ee   
An alternative form can be obtained by integrating by parts the $\partial_i$ derivative, 
leading to   
  \be\label{finalFinalb}
  \begin{split}
  \tilde{\delta} b_{ij} \ = \ & \  \partial_i\tilde{\epsilon}_j-\partial_j\tilde{\epsilon}_i  -\partial^{k}\epsilon^{l}\,\partial^{}_{[i}\,\omega_{j]kl}^{(1)}\,,  \\ 
  \ = \ &\  \partial_i\tilde{\epsilon}_j-\partial_j\tilde{\epsilon}_i -\f{1}{2}\,\partial^k\epsilon^l\,R^{(1)}_{ijkl} \;, 
\end{split}
 \ee 
with the linearized Riemann tensor 
 \be 
  R^{(1)}_{ijkl} \ \equiv \ 2\,\partial^{}_{[i}\,\omega_{j],kl}^{(1)}  \ = \ -2\, \partial_{[i}\partial_{[k}\,h_{l]j]}\;.
 \ee  
The form (\ref{finalFinalb}) shows that the gauge algebra trivializes at the linearized level. Indeed, 
as the linearized Riemann tensor is invariant under linearized gauge transformations, 
acting in the commutator with  
the inhomogeneous lowest-order variation 
gives zero. 
However, as we will show in the next subsection, this is only
an artifact of the linearization. Moreover,  to this order in a perturbative expansion, 
the final form of the gauge transformations, either (\ref{finaldel2}) or (\ref{finalFinalb}), 
cannot be reduced to the abelian gauge transformations by further field and/or parameter redefinitions, 
as we will now show. 

In order 
to see that this deformation is indeed non-trivial we first observe that a modified three-form field strength 
that is invariant under the non-trivial gauge transformations (\ref{finalFinalb}), or alternatively (\ref{finaldel2}),  is given by 
 \be\label{3formdef}
  \widehat{H}_{ijk}\, \big(b, \omega^{(1)} \big)   
   \ \equiv \ 3\Big(\partial_{[i}\,b_{jk]}-\,\omega_{[i}^{\, (1)pq}\,\partial^{}_{j}\omega^{(1)}_{k]pq}\Big)\;.   
 \ee 
Gauge invariance under (\ref{finaldel2}) 
can be easily verified using the gauge 
transformation of the linearized spin connection, $\delta \omega^{(1)}_{i,jk}=-\partial_i\partial_{[j} \epsilon_{k]}$, 
and recalling the Bianchi identity $\partial^{}_{[i}R^{(1)}_{jk]pq}=0$. 
Crucially, the modified three-form curvature is not closed but rather satisfies 
 \be\label{nontrivBianchi}
  \partial_{[i}\widehat{H}_{jkl]} \ = \ -\tfrac{3}{4}\,R^{(1)}_{[ij}{}^{pq}\,R^{(1)}_{kl]pq}\;.
 \ee  
This proves that the deformation (\ref{finaldel1}) of the gauge transformation on $b$ is non-trivial:   
a trivial deformation representing 
a field redefinition of $b$ would lead to a gauge invariant field strength
that is closed. The modification of the three-form field strength in (\ref{3formdef}), with its non-trivial Bianchi identity  
(\ref{nontrivBianchi}), is of course well-known from the Green-Schwarz mechanism for anomaly cancellation 
in heterotic string theory. 
We will show in the next subsection that its non-linear version, viewed as a deformation of the diffeomorphisms, 
defines a closed algebra with field-independent structure constants that will be related to the   
deformed Courant bracket.

\subsection{Non-linear realization and Green-Schwarz mechanism}

We start by recalling the standard formulation of the modified three-form curvature. 
This is written in terms of the Chern-Simons three-form $\Omega$
of the (Lorentz) spin connection $\omega$,  
 \be\label{furthermodH0}
   \widehat{H}_{ijk}  (b, \omega)  
   \ \equiv \ 3\Big(\partial_{[i}\,b_{jk]}+\,\Omega (\omega)_{ijk}\Big)\;,   
 \ee
where 
 \be
 \Omega(\omega)_{ijk}  \ = \ \omega_{[i}{}^{a}{}_{b}\,\partial_j\omega_{k]}{}^{b}{}_{a}
   +\f{2}{3}\,\omega_{[i}{}^{a}{}_{b}\,\omega_{j}{}^{b}{}_{c}\,\omega_{k]}{}^{c}{}_{a}\;, 
 \ee
and the spin connection $\omega_{m}{}^{ab}= -\omega_{m}{}^{ba}$  
 determined in terms of the vielbein.    Using forms, matrix notation, and 
 traces for the flat indices we have:
 \be
  \widehat H (b, \omega) 
  \ = \  {\rm d}b + \f{1}{2} \,\Omega (\omega) \,, \qquad   \Omega(\omega) \,  = \,  \,  \hbox{tr} \big(  \omega \wedge d\omega 
 + \f{2}{3}  \omega \wedge \omega \wedge \omega \bigr) \,,
 \ee
 where we define  
 \be
 \widehat H  \ \equiv \ \f{1}{3!} \, \widehat H_{ijk}\,  dx^i\wedge dx^j\wedge dx^k 
 \, , \quad  b   \ \equiv \ \f{1}{2} \, b_{ij} \, dx^i\wedge dx^j \, , \quad
 \Omega   \ \equiv \ \Omega_{ijk}\,  dx^i\wedge dx^j\wedge dx^k \,. 
 \ee
Under local Lorentz transformations with parameters $\Lambda^{ab} = - \Lambda^{ba}$, a vector $V$ transforms as   $\delta_\Lambda V^a  = -\Lambda^a{}_b V^b$, where
flat indices are raised and lowered with the Minkowski metric.  
The spin connection then transforms as 
\be 
 \delta_{\Lambda}\omega_{m}{}^{ab}\ = \ D_{m}\Lambda^{ab} \ \equiv \ \partial_m\Lambda^{ab}
 +\omega_{m}{}^{a}{}_{c}\,\Lambda^{cb}+\omega_{m}{}^{b}{}_{c}\,\Lambda^{ac}\;, 
\ee 
or, in matrix and form notation,   
\be
\delta_\Lambda \omega = {\rm d} \Lambda  + \omega \,\Lambda  -  \Lambda \, \omega\, .
\ee    
The Chern-Simons three-form varies into $\delta_\Lambda \Omega = 
\hbox{tr} ({\rm d \Lambda} \wedge {\rm d} \omega)$, which is an exact form: 
 \be
\delta_\Lambda \Omega \ = 
\ - {\rm d}\,  \hbox{tr} ( {\rm d}\Lambda \wedge \omega\, )   \,. 
 \ee  
From this transformation behavior it follows that  $\widehat H (b, \omega) $ 
can be made gauge invariant 
by assigning to $b$ a non-standard variation under local Lorentz transformations, 
 \be\label{nonlinLor}
 \delta_\Lambda b \ = \ \f{1}{2}  \,\hbox{tr} \, ( {\rm d} \Lambda \wedge\omega\, ) \quad 
 \to \quad  \delta_{\Lambda}b_{ij} \ = \  -\,  \partial_{\,[i}\, \Lambda^{ab}\,\omega_{j]ab}\;.  
 \ee
This is the transformation needed for Green-Schwarz anomaly cancellation.  

In the above we have deformed the local Lorentz transformations 
by assigning a non-trivial transformation to the Lorentz singlet $b_{ij}$, but left the 
action of the diffeomorphisms unchanged (the 
Lorentz Chern-Simons term is a three-form under diffeomorphisms). 
Consequently, the diffeomorphism algebra is unaffected, but rather the Lorentz gauge algebra 
becomes non-trivial. An explicit computation with (\ref{nonlinLor}) shows that  
the deformed local Lorentz transformations close on $b_{ij}$ as 
 \be\label{LORclosure}
  \big[\delta_{\Lambda_1},\delta_{\Lambda_2}\big]b \ = \ 
  \delta_{[\Lambda_1,\Lambda_2]}\,b\, +\,{\rm d}\tilde{\xi}_{12}\;, 
  \ee 
with the usual commutator of two Lorentz transformations and an extra one-form
$\xi_{12} =  \xi_{12i} \, dx^i$ given by  
 \be
  \tilde{\xi}_{12} \ = \ -\f{1}{2} \hbox{tr} ( \Lambda_1 \,{\rm d} \Lambda_2 -   
  \Lambda_2 \,{\rm d} \Lambda_1)  \quad \to \quad 
  \tilde{\xi}_{12\, i} \ = \ \f{1}{2}\,\bigl( \Lambda_{1}^{ab}\,\partial_i\Lambda_{2ab}-\Lambda_{2}^{ab}\,\partial_i\Lambda_{1ab} \bigr)\;. 
 \ee  
The gauge algebra is 
field-independent.\footnote{Had we   
chosen the equally valid
$b$ field transformation $\delta_\Lambda b  =  -\f{1}{2}  \,\hbox{tr} \, ( \Lambda {\rm d} \omega\, )$, the result would have been an algebra with field-dependent structure constants.  }

In order to make contact with the 
deformed Courant bracket 
we present now an equivalent form of the gauge transformation on $b_{mn}$
that deforms the diffeomorphisms rather than the local Lorentz transformations. 
Thus, here we use a modification of the three-form curvature by a Chern-Simons form based on the 
Christoffel symbols rather than the spin connection, i.e., 
 \be\label{furthermodH}
   \widehat{H}_{ijk} \big(b, \Gamma\big) 
   \ \equiv \ 3\Big(\partial_{[i}\,b_{jk]} +\Omega(\Gamma)_{ijk}\Big)\;, 
 \ee
where 
 \be
  \Omega(\Gamma)_{ijk} \ = \ \Gamma_{[i |p|}^{\,q}\partial^{}_j\Gamma_{k]q}^{\,p}+\f{2}{3}
 \, \Gamma_{[i|p|}^{\,q}\Gamma_{j|r|}^{\,p}\Gamma_{k]q}^{\,r}\;. 
 \ee
 In the language of forms and matrices we have   
 \be
  \widehat H (b, \Gamma)  
  \ = \  {\rm d}b + \f{1}{2} \,\Omega (\Gamma) \,, \qquad   \Omega(\Gamma) \,  = \,  \,  \hbox{tr} \big(  \Gamma \wedge {\rm d}\Gamma 
 + \f{2}{3}\,  \Gamma \wedge \Gamma \wedge \Gamma \bigr) \,,
 \ee
 where we define the matrix valued one-forms  $\Gamma$ 
 as well as the matrix representation of the Christoffel symbols 
 \be
(\Gamma )^k{}_l \ \equiv  \  (\Gamma_i)^k{}_l \, dx^i \ \equiv \  \Gamma_{\, il}^{\,k} \, dx^i  
 \, .  \ee
The Christoffel symbols are determined in terms of the metric by 
 \be\label{Christoffel}
  \Gamma_{mn}^{\,k} \ = \ \frac{1}{2}g^{kl}\big(\partial_m g_{nl}+\partial_n g_{ml}-\partial_l g_{mn}\big)\;,    
 \ee 
and transform under diffeomorphisms as 
\be
\delta_{\xi}\Gamma_{mn}^{\,k} \ = \ 
{\cal L}_{\xi}\Gamma_{mn}^{\,k}+\partial_m\partial_n\xi^k\,.
\ee  
It is convenient, for general objects $A$, 
to write $\delta_\xi  A = {\cal L}_{\xi} A + \Delta_\xi A$, where $\Delta_\xi A$ denotes the failure of $A$ to transform as
a tensor.  In this notation $\Delta_\xi \Gamma_{mn}^{\,k}= \partial_m\partial_n\xi^k$, 
which we can write as
\be
\Delta_\xi \Gamma \ = \ {\rm d}  (\partial \xi) \,,
\ee
where we used the matrix notation $(\partial \xi)^k{}_n \equiv  \partial_n \xi^k$. 
One may also verify that 
\be
\Delta_\xi  {\rm d}\Gamma\ = \ - \Gamma \wedge {\rm d}  (\partial \xi)  - {\rm d}  (\partial \xi) \wedge \Gamma \,. 
\ee
With the help of the last two equations it is straightforward to show that the failure of 
the Chern-Simons form $\Omega(\Gamma)$  to be a tensor is an exact three-form:
\be
\Delta_\xi  \Omega (\Gamma)  \ = \  \hbox{tr} \bigl(  {\rm d}(\partial \xi) \wedge {\rm d} \Gamma\bigr)  \ = \ {\rm d}\,  \hbox{tr}\bigl( (\partial \xi)  {\rm d} \Gamma\bigr) \ = \ 
 {\rm d} \, \hbox{tr}\bigl( -{\rm d}(\partial \xi)  \wedge \Gamma\bigr) \,.
\ee
Again, we can assign a suitable   
transformation  $\Delta_\xi b$  
 so that the curvature $\widehat H (b, \Gamma)$ 
is diffeomorphism covariant:  $\Delta_\xi \widehat H =0$.  
The two ways of writing $\Delta_\xi \Omega$ as an exact form give us two options:
\be
\label{twovmoptions}
\Delta_\xi b \ = \ -\f{1}{2} \, \hbox{tr} (  \partial \xi  \, {\rm d} \Gamma \, )  \,, \qquad \hbox{or}
\qquad  \Delta_\xi b \ = \ \f{1}{2} \, \hbox{tr}  \bigl(  {\rm d}(\partial \xi)\wedge \Gamma \, \bigr) \,. 
\ee
At this point we can try to consider which option gives
a non-linear completion of 
(\ref{finalFinalb}).\footnote{The  
 choice to replace $R^{(1)}$ 
by the full Riemann tensor does not lead
to the correct  
result.}  In component notation, the first option gives 
 \be\label{nonlinDiff}
  \delta_{\xi}b_{ij} \ = \ {\cal L}_{\xi} b_{ij}-\partial_p\xi^q\,\partial^{}_{[i}\,\Gamma_{j]q}^{\,p}\;. 
 \ee
One may verify, using (\ref{Christoffel}), that this expression reduces to 
(\ref{finalFinalb}) upon
 expansion about flat space with $g_{ij}=\eta_{ij}+h_{ij}$.  
This transformation actually  gives a gauge algebra with
field-dependent structure constants.  The second option in (\ref{twovmoptions})
is the analog of (\ref{nonlinLor}), and gives
 \be\label{nonlinDiff2}
  \delta_{\xi+\tilde{\xi}}\, b_{ij} \ = \ 2\, \partial_{[i}\, \tilde{\xi}_{j]} 
  + {\cal L}_{\xi} b_{ij} \ 
  + \ \partial^{}_{[i}\partial_{p}\xi^q\; \Gamma_{j]q}^{\, p}\; ,  
 \ee
 or, in form notation, 
 \be
 \label{nonlinDiff2v2}
 \delta_{\xi+\tilde{\xi}}\, b \ = \ {\rm d} \tilde{\xi} 
 \,  +\,  {\cal L}_{\xi} b \ + \ \f{1}{2} \, \hbox{tr} \bigl(  {\rm d} (\partial \xi) \wedge \Gamma \bigr) \,.
 \ee
The gauge algebra based on (\ref{nonlinDiff2v2})
is field-independent and can be directly related to the deformed Courant bracket to be discussed below.  
Indeed, a direct computation of the gauge algebra with (\ref{nonlinDiff2v2})
quickly yields 
  \be
 \begin{split}
 \big[\,\delta_{\xi_1+\tilde{\xi}_1}\,,\;\delta_{\xi_2+\tilde{\xi}_2}\,\big]\,  b
 \ \ =  & \ \ \ {\cal L}_{\xi_2}  {\rm d} \tilde \xi_1  -  {\cal L}_{\xi_1}  {\rm d} \tilde \xi_2 
 \, 
 - \f{1}{2} {\rm d} \, \hbox{tr} \bigl(    {\rm d} (\partial \xi_2)\partial {\xi_1} - 
   {\rm d} (\partial \xi_1)\partial {\xi_2}\bigr) \\
 &  + \,  {\cal L}_{[\xi_2, \xi_1] } b\, +  \f{1}{2} \hbox{tr} \bigl( \bigl[  {\cal L}_{\xi_2}  {\rm d} (\partial \xi_1) - {\cal L}_{\xi_1}  {\rm d} (\partial \xi_2)\bigr] \wedge \Gamma\bigr) \,.
 \end{split}
 \ee
 Noting that exterior derivatives and Lie derivatives commute allows to simplify the first line, and another short calculation allows one to simplify the second term on the second line.  The result is
 \begin{equation*}
 \begin{split}
 \hskip-10pt\big[\,\delta_{\xi_1+\tilde{\xi}_1}\,,\;\delta_{\xi_2+\tilde{\xi}_2}\,\big] b
 \  =  & \ \ {\rm d} \Bigl( {\cal L}_{\xi_2}  \tilde \xi_1  -  {\cal L}_{\xi_1} \tilde \xi_2 
 \,  -\f{1}{2} {\rm d}\big(i_{\xi_2}\tilde{\xi}_1-i_{\xi_1}\tilde{\xi}_2\big) 
 - \f{1}{2}  \, \hbox{tr} \bigl(    {\rm d} (\partial \xi_2)\partial {\xi_1} - 
   {\rm d} (\partial \xi_1)\partial {\xi_2}\bigr) \Bigr) \\
 &  + \,  {\cal L}_{[\xi_2, \xi_1] } b\, +  \f{1}{2} {\rm d} \bigl( \partial [\xi_2, \xi_1] \bigr) 
  \wedge \Gamma \,.
 \end{split}
 \end{equation*}
We see that the right-hand side takes the form of a gauge transformation of $b$ as
in (\ref{nonlinDiff2v2}).   The vector parameter of the resulting transformation
is $[\xi_2, \xi_1]$, which is the vector part of $[\xi_2 + \tilde \xi_2, \xi_1 + \tilde \xi_2]'$.  The one-form parameter is that in parenthesis on the 
first line of the above equation.  It is indeed equal to the one-form part of 
$[\xi_2 + \tilde \xi_2, \xi_1 + \tilde \xi_2]'$ as one can confirm comparing with 
(\ref{CourantIntro}).   All in all we have proven that 
 \be
  \big[\,\delta_{\xi_1+\tilde{\xi}_1}\,,\;\delta_{\xi_2+\tilde{\xi}_2}\,\big] b 
  \ = \ 
  \delta_{\big[\xi_2+\tilde{\xi}_2\,,\;\xi_1+\tilde{\xi}_1\big]'}\,b\;. 
 \ee
This shows that the gauge transformations (\ref{nonlinDiff2}) provide an exact realization of the 
deformed Courant bracket as the gauge algebra. Moreover, as the gauge transformations of the metric 
$g_{mn}$ are undeformed it is evident from (\ref{nonlinDiff2}) that the deformation is exact in $\alpha'$. 

Some remarks are in order regarding the equivalence of the Chern-Simons forms based 
on $\omega$ and $\Gamma$, see, e.g., \cite{Guralnik:2003we}.  For this purpose first
recall that under transformations of the spin connection of the form
\be
\label{transform}
\omega' \ = \  U^{-1} {\rm d} U + U^{-1} \omega U  \,, 
\ee
the Chern-Simons form transforms as follows:
\be
\label{CS-trans}
\Omega(\omega') \ = \  \Omega (\omega)  \, - \, 
{\rm d}  \, \hbox{tr} \,\bigl(  {\rm d} U\, U^{-1} \wedge \omega  \bigr) - \f{1}{3} \, \hbox{tr} [ (U^{-1} {\rm d} U)\wedge (U^{-1} {\rm d} U)  \wedge (U^{-1} {\rm d} U) ]  \,.
\ee
When the matrix $U$ is a Lorentz transformation, this is a gauge transformation from 
$\omega$ to $\omega'$.
If the matrix $U$ is more general, $\omega'$ would not be a spin connection, but
the above still holds as an identity relating the Chern-Simons 
terms constructed from $\omega$ and $\omega'$.   One can relate in this way the spin connection to the Christoffel connection. Indeed, by
 the `vielbein postulate' these connections are related 
by 
 \be
  D_m e_{n}{}^{a} \  \equiv \  
  \partial_m e_{n}{}^{a}+\omega_{m}{}^{a}{}_{b}\,e_{n}{}^{b}-\Gamma_{mn}^{\,k}
  e_{k}{}^{a} \ = \ 0\;. 
 \ee
Recalling our matrix notation for these connections and introducing one 
more
for the vielbein and inverse vielbein, 
 \be
  ({\bf \omega}_m)^{a}{}_{b} \ \equiv \  \omega_{m}{}^{a}{}_{b}\;, \quad  
 ({\Gamma}_m)^k{}_n \ \equiv \ \Gamma_{mn}^{\,k}\;, \quad 
 ({e})^{\,a}{}_{m} \ \equiv \ e_m{}^{a}\;, \quad 
 ({ e}^{-1})^{m}{}_a \ \equiv \ e_{a}{}^{m}\;, 
\ee 
the vielbein postulate  implies that the connection one-forms are related by  
 \be
  {\Gamma} \ = \ { e}^{-1}{\rm d}\, { e}+{ e}^{-1}{ \omega} \,e\;. 
 \ee 
This relation is of the form (\ref{transform}), with $U= e$, which is not a Lorentz
transformation. It thus follows that
\be
\Omega (\Gamma) \ = \ \Omega ( \omega) - \, 
{\rm d}  \, \hbox{tr} \,\bigl(  {\rm d} e\, e^{-1} \wedge \omega  \bigr) - \f{1}{3} \, \hbox{tr} [ (e^{-1} {\rm d} e)\wedge (e^{-1} {\rm d} e)  \wedge (e^{-1} {\rm d} e) ]\,. 
\ee
We see that the two Chern-Simons forms differ by an exact two-form and a closed
three-form whose integral is associated with the winding number of the transformation
matrix.  
Therefore, at least locally the difference between the two Chern-Simons forms is exact, and
the field strengths $\widehat H (b, \omega)$ and $\widehat H(b, \Gamma)$ can
be made to agree after a field redefinition of $b$.  Thus the two formulations
can be treated as equivalent.\footnote{The same topological subtleties arise in proving the gauge invariance of $\widehat H$ under large gauge transformations.}

Given the above results, it follows that the construction of \cite{Hohm:2013jaa}
gives a manifestly and exactly 
T-duality invariant theory that incorporates the $\alpha'$
corrections of the Green-Schwarz mechanism.
Since we have a gauge invariant field strength $\widehat H$ 
the action  
  \be\label{original}
  S \ = \ \int d^Dx \sqrt{-g}e^{-2\phi}\bigl[\, R+4(\partial\phi)^2-\f{1}{12}\,\widehat{H}^{mnk}\widehat{H}_{mnk}\bigr]  \;,  
 \ee
is at least a subsector of the exactly T-duality invariant theory in \cite{Hohm:2013jaa}.
Expanding the $\widehat{H}^2$ term above 
one obtains structures with up to six derivatives, which is as predicted by the full theory 
constructed in \cite{Hohm:2013jaa}.
Most likely, when expressed in terms of $g$ and $b$, the exactly duality
invariant theory will have corrections to all orders in $\alpha'$.
We finally note that the above action 
corresponds to the truncation 
of heterotic string theory that sets the Yang-Mills gauge fields to zero. 
These gauge fields 
can be naturally included in DFT, at least for the abelian subsector, by enlarging 
$O(D,D)$ to $O(D,D+n)$, with $n$ the number of gauge vectors \cite{Siegel:1993th,Siegel:1993bj,Hohm:2011ex}.
(See also \cite{Baraglia,Garcia-Fernandez,Anderson:2014xha,delaOssa:2014cia} 
for Courant algebroids in `generalized geometry' formulations of heterotic strings.)

\section{Courant bracket, C-bracket and their automorphisms} 

In this section we review the $B$ automorphism of the Courant bracket and
find its extension to the C-bracket.  
This automorphism is preserved by the deformation discussed in the next section. 
The  Courant bracket for elements $V+\tilde{V}\in T\oplus T^*$,  where $V$ is 
a vector and $\tilde V$ a one-form, takes the form 
 \be\label{classCourant}
  \big[\, V+\tilde{V}, W+\tilde{W} \,\big]  \ = \ \big[V,W]+{\cal L}_{V}\tilde{W}-{\cal L}_{W}\tilde{V}
  -\f{1}{2} {\rm d}\big(i_{V}\tilde{W}-i_W\tilde{V}\big)\; .
 \ee 
Here $[V, W]$ is the Lie bracket of vector fields, ${\cal L}$ denote Lie derivatives,
and $i_{V}\tilde{W}=V^i\tilde{W}_i$ for a vector $V = V^i \partial_i$ and a
one-form $\tilde W = \tilde W_i dx^i$. 
The last term on the right-hand side is an exact one-form. 
Its coefficient is fixed by the condition that the bracket have an extra automorphism parameterized by an arbitrary
closed two-form $B$:
 \be\label{automorphism}
B \hbox{ transformation:} \quad   V+\tilde{V}\;\rightarrow\; V+\tilde{V}+i_VB\;,   \qquad  {\rm d}B \ = \ 0\;. 
 \ee
 Here $i_V B$ is the one-form obtained by contraction:  $(i_V B) (W) =  B (V, W)$ or, more
 explicitly, $i_VB =  V^i B_{ij} \, dx^j$ when $B = \f{1}{2} B_{ij}\, dx^i \wedge dx^j$ (we use $dx^i \wedge dx^j = dx^i\otimes dx^j - dx^j\otimes dx^i)$.     
Under a $B$ transformation the one-form part of an element of the algebra
 is shifted as $\tilde{V}_i\,\rightarrow\, \tilde{V}_i+V^j B_{ji}$.
The statement that a $B$ transformation is an automorphism of the bracket means that
  \be
    \big[\, V+\tilde{V}+i_VB\, , \, W+\tilde{W}+i_WB \,\big] \ = \ \big[\, V+\tilde{V}\, , \,
    W+\tilde{W} \,\big]\,
    +\, i_{[V,W]} B \;. 
 \ee 
This property is readily checked using the identities ${\cal L}_V = i_V {\rm d} + {\rm d}\,  i_V$, 
$[{\cal L}_V , i_W] = i_{[V, W]}$ and $i_V i_W = - i_W i_V$. 

In the doubled geometry 
we have now generalized vectors $V^M (X)$ or $\xi^M(X)$ 
on a suitably generalized doubled manifold with coordinates $X^M$,  $M = 1, \ldots, 2D$.
These vectors and partial derivatives are 
decomposed as  
\be
V^M \ = \ \begin{pmatrix} \tilde V_i \\ V^i \end{pmatrix} \,, \  
\qquad  \xi^M \ = \   \begin{pmatrix} \tilde \xi_i \\ \xi^i \end{pmatrix}   \,, \qquad
  \partial_M \ = \  
\begin{pmatrix} \tilde \partial_i \\ \partial_i \end{pmatrix} \,,
\ee 
with $\partial_i = {\partial \over \partial x^i}$ and $\tilde \partial^i = {\partial \over \partial \tilde x_i}$.  Generalized Lie derivatives are defined by
 \be
 \label{gen-Lie-der}
  \widehat{\cal L}_{\xi}V^M \ = \ \xi^N\partial_N V^M+\big(\partial^M\xi_N-\partial_N\xi^M\big)V^N\;, 
 \ee
where indices are raised and lowered with the constant $O(D,D)$ metric
$\eta_{MN}$ and its inverse $\eta^{MN}$.   They define a generalized notion of diffeomorphisms. 
For objects with additional 
indices the generalized Lie derivative includes extra terms.  
Objects that transform under generalized diffeomorphisms with these 
generalized Lie derivatives are called generalized tensors.  
The  C-bracket $[ \ , \ ]_{{}_C}$  is defined by 
\be
\bigl[ \,  V \, , \,  W \, \bigr]_{{}_{\rm{C}}}^M  \ = \ [ \,  V \, , \, W \, ]^M   -  \, \f{1}{2} \,\bigl( \,
 V^P \partial^M W_P  -  
W^P \partial^M V_P\, \bigr)  \,, 
\ee
where $[ V, W]^M \equiv  V^K\partial_K W^M - W^K \partial_K V^M$ 
is the
analog of the Lie bracket.  When we choose a section, say $\tilde \partial^i =0$,
the C-bracket reduces to the Courant bracket.  

We now ask: What is the automorphism of the C-bracket that corresponds
to the $B$-transformation of the Courant bracket?
In analogy to the earlier analysis we consider the transformation
induced by an antisymmetric two-index generalized tensor $B^{MN} = - B^{NM}$:
 \be\label{ODDShift}
  V^M\,\rightarrow \, V^M-B^{MN} V_N\,,  \quad \hbox{or} \quad  V \,\rightarrow \, V-B V\;. 
 \ee
Note that this transformation,
infinitesimally, can be viewed as a {\em local }  $O(D,D)$ transformation.
As such,  it is somewhat surprising that it can be an invariance of the theory.
The automorphism would require that 
\be
\label{B-auto-C-bracket}
  \big[\, V-BV\,,\, W-BW\,\big]^M_{{}_{\rm{C}}} \ = \ 
  \big[V,W\big]^M_{{}_{\rm{C}}}-B^{MN}\big[V,W\big]_{{}_{\rm{C}}}{}_N\;.
 \ee 
A short calculation shows that 
\be
\label{vmvm}
\begin{split}
 \hskip-5pt \big[\, V-BV\,,\, W-BW\,\big]^M_{{}_{\rm{C}}} \ = \ & \  
  \big[V,W\big]^M_{{}_{\rm{C}}}-B^{MN}\big[V,W\big]_N  \\[0.5ex]
  & \hskip-10pt 
  - (\partial^K B^{MP} + \partial^P B^{KM} + \partial^M B^{PK} )  V_K W_P  \\ 
 & \hskip-10pt  -(BV)^K \partial_K (W - BW)^M   + \f{1}{2} (BV)^P \partial^M (BW)_P -
 (V\leftrightarrow W) \,.  
 \end{split}
\ee
It is now natural to demand, in analogy to the condition ${\rm d}B =0$ 
for the $B$ automorphism of the Courant bracket, that 
  \be\label{Cond1}
   \partial^{[M} B^{NK]} \ = \ 0\;. 
  \ee
This eliminates the second line in (\ref{vmvm}), but this is not sufficient for the
autormorphism to hold.  There remain two problems.  First, the bracket on the last term
of the first line is a Lie bracket, not a C-bracket, as required for the automorphism.
Second,   the terms on the last line do not cancel.  These difficulties are related and
one clue is the fact that the condition (\ref{Cond1}) is not covariant under generalized
diffeomorphisms.  Denoting by $\Delta_\xi$ the failure of an object to be
a generalized tensor, one quickly finds that 
 \be
  \Delta_{\xi}\big(\partial^{[M}B^{NK]}\big) \ \equiv \  \partial^{[M} \widehat {\cal L}_\xi B^{NK]} - \widehat {\cal L}_\xi \big(\partial^{[M}B^{NK]}\big) \ = \ 
  2\ \partial_P\partial^{[M}\xi^{N} B^{K]P}\;. 
 \ee 
This is zero if we demand that $B^{MN}$ is `covariantly constrained' (a notion introduced in \cite{Hohm:2013jma}) 
in the sense that 
derivatives along $B$ vanish: 
 \be\label{Cond2}
  B^{MN} \partial_N \ = \ 0\;.
 \ee
This condition helps in two ways.  First we have that
\be
B^{MN}\big[V,W\big]_N\ = \ 
B^{MN}\big[V,W\big]_{{}_{\rm{C}}}{}_N\,, 
\ee
since the extra term in the C-bracket has a derivative tied to the $B$ field.
Moreover, the first term on the third line of (\ref{vmvm}) vanishes. Therefore,
so far we have 
\be
\label{vmvmvm}
\begin{split}
 \hskip-5pt \big[\, V-BV\,,\, W-BW\,\big]^M_{{}_{\rm{C}}} \ = \ &   \  
  \big[V,W\big]^M_{{}_{\rm{C}}}-B^{MN}\big[V,W\big]_{{}_{\rm{C}}}{}_N\\[0.1ex]
  & + \f{1}{2} (BV)^P \partial^M (BW)_P -
 (V\leftrightarrow W) \,.  
 \end{split}
\ee
The second line should still vanish.  We can understand how this happens
by looking in detail at the constraint $B^{MN} \partial_N=0$:
\be
\begin{split}
B^{ij} \partial_j + B^i{}_j \tilde \partial^j \ =\ & \ 0 \,, \\
B_i{}^{j} \partial_j + B_{ij}  \tilde \partial^j \ =\ & \ 0\,. 
\end{split}
\ee
Solving the strong constraint by declaring $\tilde \partial =0$ these conditions become
\be
B^{ij} \partial_j  \ = \ 0  \,, \qquad 
B_i{}^{j} \partial_j  \ = \   0\,.
\ee
For these conditions to hold in all generality (namely, for brackets of arbitrary
elements) we must set $B^{ij}$ and $B_i{}^j = - B^j{}_i$ equal to zero,  and the only surviving component
of $B^{MN}$ is  $B_{ij}$:
\be
B^{ij} \ = \ 0 \, , \quad   B_i{}^j \ = \ - B^j{}_i \ = \ 0 \,,   \qquad  B_{ij} \;\; \hbox{nonzero}\,. 
\ee
This is the general solution of $B^{MN}\partial_N=0$. At this point, 
(\ref{Cond1}) requires  that $B_{ij}$ is a closed two-form.  
This is consistent with the $B$ automorphism
of the Courant bracket, which should arise upon reduction to the non-doubled
space.  We now note that if $B^{MN}$ has only components $B_{ij}$,
then any contraction of indices between two $B$ fields must vanish
\be
\label{Cond3}
{\cal O} B^{MN} {\cal O}'  B_{MK}  \ = \ 0 \,, 
\ee
where ${\cal O}$ and ${\cal O}'$ denote arbitrary factors that may include derivatives.  The second line
in (\ref{vmvmvm}) features such a contraction. Those terms thus vanish, showing 
that we have the $B$ automorphism.  
In summary, the C-bracket has the automorphism 
\be
V^M\,\rightarrow \, V^M-B^{MN} V_N \,, 
\ee
when the $B$ field satisfies 
\be\label{covconstrained}
B^{MN} \ = \ - B^{NM} \,, \quad  \partial^{[M} B^{NK]}  \ = \ 0 \,,  \quad
 B^{MN} \partial_N \ = \ 0\,,\quad {\cal O} B^{MN} {\cal O}'  B_{MK}  \ = \ 0\,.
\ee
Although, as noted above, (\ref{Cond3}) identically holds if the condition $B^{MN} \partial_N = 0$ is solved, 
this does not seem derivable in an $O(D,D)$ covariant way, and so here we included the last condition. 

To understand better the automorphism, we consider the  familiar
automorphism of the C-bracket generated by generalized Lie derivatives. 
Indeed, for infinitesimal parameters $\lambda$ we have that $V \to V + \lambda 
\widehat{\cal L}_\xi V$ is an automorphism since
we have 
 \be
  \widehat{\cal L}_{\xi}\,\big[\, V,W\big]_{{}_{\rm{C}}} \ = \ \big[ \,\widehat{\cal L}_{\xi}V\, ,\, W\big]_{{}_{\rm{C}}}+\big[\, V\,, \, \widehat{\cal L}_{\xi}W\big]_{{}_{\rm{C}}}\;.
 \ee  
 The finite version of this automorphism holds for exponentials of generalized Lie derivatives,  
 \be
  e^{\widehat {\cal L}_\xi} \big[\, V,W\big]_{{}_{\rm{C}}} \ = \ \big[ \,e^{\widehat {\cal L}_\xi}V\, ,\, e^{\widehat {\cal L}_\xi} W\big]_{{}_{\rm{C}}} \,.
  \ee  
We now argue that, at least locally, we can view an infinitesimal $B$ transformation
as generated by a Lie derivative.  The condition $\partial_{[M} B_{NK]} = 0$ implies
that locally there exists a $\xi^M$ such that
\be
B_{MN} \ = \ \partial_M \xi_N - \partial_N \xi_M\,.
\ee  
Since $B_{MN}$ needs to satisfy
$B^{MN}\partial_N = 0$, we demand,  in addition, that $\xi^K \partial_K = 0$.
 (This is clear in the  frame $\tilde \partial=0$, 
 since only $B_{ij}$ exists 
 and thus we can set $\xi^i=0$ resulting in $\xi^K \partial_K = 0$.)
As a result, a generalized Lie derivative along $\xi$ indeed
amounts to a $B$ transformation:
 \be
  \widehat{\cal L}_{\xi}V_M \ = \ \xi^K\partial_K V_M+\big(\partial_M\xi_N-\partial_N\xi_M\big)V^N \ = \ 
  B_{MN} V^N  \;.  
 \ee
The generalization of the above discussion to the global aspects of a doubled (generalized) manifold may be
of interest. 

We conclude this section with a simple observation that explains why 
the reduction of generalized Lie derivatives of the doubled theory give
automorphisms of the Courant bracket. 
Consider the expression
(\ref{gen-Lie-der}) from the doubled geometry and set
$\tilde{\partial}=0$.  We then find that the generalized Lie derivative of 
$V^M=(\tilde{V}_i,V^i)$ reads
 \be
 \begin{split}
  \widehat{\cal L}_{\xi} {V} \ = \ & \  {\cal L}_\xi V \,, \\ 
   \widehat{\cal L}_{\xi}\tilde{V}_i \ = \ &
  \  {\cal L}_\xi \tilde V_i  \, +\,  \big( \p_i \tilde \xi_j - \p_j \tilde \xi_i\big) V^j\;. 
\end{split}
 \ee
 The last term in the second equation can be written as a $B$-transformation: 
 \be
   \widehat{\cal L}_{\xi}\tilde{V} \ = \ 
  \  {\cal L}_\xi \tilde V  \, +\,  i_V B \, , \qquad  
   B = -\tfrac{1}{2}( \p_i \tilde \xi_j - \p_j \tilde \xi_i\big)\,  dx^i \wedge dx^j \;. 
 \ee
Ordinary Lie derivatives, of course, generate autormorphisms of the Courant bracket.
The generalized Lie derivatives that
arise from the doubled theory are 
automorphisms as well, because the extra terms
beyond ordinary Lie derivatives
are $B$ automorphisms.

\section{Exact deformation of the Courant bracket}
We now turn to the deformation of the C-bracket introduced in \cite{Hohm:2013jaa}, which reads 
 \be\label{defbracket}  
  \big[V,W\big]'^M \ = \ \big[V,W\big]_{{}_{\rm C}}^M+\f{1}{2}\big(\partial_KV^L\partial^M\partial_LW^K
  -(V\leftrightarrow W)\big)\;. 
 \ee 
Let us discuss a few of its properties. First,  
because of the constraint  (\ref{Cond2}), the $B$ transformation (\ref{ODDShift})
is also an automorphism of the deformed C-bracket. 
Indeed,  
 \be\label{defbracketBauto}  
  \big[V-BV,W-BW\big]'^M \, = \, \big[V-BV,W-BW\big]_{{}_{\rm C}}^M+\f{1}{2}\big(\partial_KV^L\partial^M\partial_LW^K
  -(V\leftrightarrow W)\big)\;, 
 \ee 
because the vector fields $V$ and $W$ in the extra term are contracted with
derivatives and thus the $B$ shift drops out.  Since the C-bracket has the
$B$ automorphism (\ref{B-auto-C-bracket})
 \be\label{defbracketBauto99}
 \begin{split}  
  \hskip-10pt \big[V-BV,W-BW\big]'^M \, = \, & \ \big[V,W\big]^M_{{}_{\rm{C}}}-B^{MN}\big[V,W\big]_{{}_{\rm{C}}}{}_N +\f{1}{2}\big(\partial_KV^L\partial^M\partial_LW^K
  -(V\leftrightarrow W)\big) \\
  \, = \, & \ \big[V,W\big]'^M-B^{MN}\big[V,W\big]_{{}_{\rm{C}}}{}_N \,, \\
    \, = \, & \ \big[V,W\big]'^M-B^{MN}\big[V,W\big]'_N \,, 
\end{split}
 \ee 
where the last substitution is allowed because the vector index in the correction
of the C-bracket is carried by a derivative. 
The  $B$ automorphism is thus unchanged.

The deformed C-bracket can be realized  
as the gauge algebra for deformed generalized Lie 
derivatives. On a vector these transformations read 
 \be\label{genLie}
  \delta'_{\xi}V^M \ =    
    \ {\bf L}_{\xi}V^M \ \equiv \ \widehat{\cal L}_{\xi}V^M-\partial^M\partial_K\xi^L \partial_L V^K\;, 
 \ee  
which close according to (\ref{defbracket}). There is also a deformation of the inner product defined 
by the $O(D,D)$ invariant metric, 
 \be\label{deformedInner}   
 \langle V| W\rangle'  \ \equiv \ \langle V| W\rangle-\partial_MV^N\partial_NW^M
 \ = \ \ V^M W^N\eta_{MN}-\partial_MV^N\partial_NW^M\;.
\ee
Indeed, it is straightforward to verify that this transforms as a scalar under (\ref{genLie}). 
 
In the remainder of this section we investigate these deformed structures on the physical $D$-dimensional 
subspace and show how they provide a consistent non-trivial deformation of the Courant bracket of 
generalized geometry.  
Setting $\tilde{\partial}^i=0$, one finds for the ${\rm C}'$-bracket that the vector part is not corrected,
but the one-form part is, 
\be\label{DEfCbracket}
\begin{split}
\big[ V , W\big]'^i  \ = \ & \ \big[V,W\big]^i   \,, \\[0.5ex]
 \big[ V , W\big]'_i  \ = \ &  \ \big[ V , W\big]_{{\rm C}\;i}   
\ - \  {\textstyle {1\over 2}} \, \partial_i \partial_\ell V^k  \,\partial_k W^\ell
\   + \  {\textstyle {1\over 2}} \, \partial_i \partial_\ell W^k\,\partial_k V^\ell \,.
\end{split}
\ee
Similarly, the deformed generalized Lie derivative on the vector part is not corrected but on the 
one-form part it is,  
 \be
\label{corrliegen}
\begin{split}
({\bf L}_\xi  V)^i   \ = \ & \ \xi^k \p_k V^i - V^k \p_k \xi^i \;,   \\
({\bf L}_\xi  \tilde{V})_i  \ = \ & \ \xi^k \p_k \tilde{V}_i\, + \, \p_i \xi^k \,\tilde{V}_k  
+ (\p_i \tilde \xi_k - \p_k \tilde \xi_i) V^k \, - \, \p_i \p_k \xi^l \p_l V^k \,.
\end{split}
\ee
The deformed inner product (\ref{deformedInner}) reads 
\be\label{INNERMORE}
\langle V+\tilde{V} | W+\tilde{W}\rangle'   \ = \
\langle V+\tilde{V} | W+\tilde{W}\rangle   \ - \  \p_i V^j   \p_j W^i\ = \  V^i \tilde{V}_{i} + W^i \tilde{W}_{i}  \ - \  \p_i V^j   \p_j W^i \,. 
\ee
The deformed Courant bracket has a non-trivial Jacobiator that is, however, exact. 
Specifically, the Jacobiator 
 \be
  J_{C'}(U+\tilde{U},V+\tilde{V},W+\tilde{W}) \ \equiv \ \sum_{{\rm cycl}}\big[\big[U+\tilde{U},V+\tilde{V}\big]',W+\tilde{W}\big]'\;,
 \ee 
where the cyclic sum has three terms with coefficient $1$, 
reads 
 \be
   J_{C'}(U+\tilde{U},V+\tilde{V},W+\tilde{W}) \ = \ \frac{1}{6}
   \,{\rm d}\,\Big(\sum_{{\rm cycl}}\Big\langle \big[U+\tilde{U},V+\tilde{V}\big]',W+\tilde{W}\Big\rangle' 
   \Big)\;. 
  \ee 
The Jacobiator takes a form fully analogous to that of the undeformed Courant bracket, but 
with the bracket and inner product replaced by the deformed bracket and inner product. 
This result follows immediately from the proof given \cite{Hohm:2013jaa} for the 
deformed 
C-bracket. 
The deformations above are the full deformations (no higher orders in $\alpha'$ are 
needed) 
and are mutually compatible in that the deformed 
bracket transforms covariantly under the deformed generalized Lie derivatives, etc. 
In the following we establish this in some detail in order to elucidate more the novel geometrical 
structures.

We start by introducing some useful index-free notation. For the partial derivative of a vector $V$ we 
use a matrix notation and, moreover, if we want to stress the interpretation of $V$ as a differential 
operator we put a vector arrow on top, 
 \be
  \partial V \ \equiv \ \big(\partial_i V^j\big)\;, \qquad \vec{V} \ \equiv \ V^i\partial_i\;. 
 \ee
The partial derivative  $\partial V$ is not a tensor of type $(1,1)$.   
Rather, it has an 
anomalous transformation under infinitesimal diffeomorphisms $\delta_{\xi}V\equiv {\cal L}_{\xi}V$ generated by Lie derivatives.  
Indeed, a quick computation in local coordinates gives 
 \be\label{ANOmalous}
  \delta_{\xi}\big(\partial_iV^j\big) \ = \ {\cal L}_{\xi}\big(\partial_iV^j\big)-V^k\partial_k\partial_i\xi^j \;. 
 \ee 
Here, with 
slight abuse of notation, we mean that ${\cal L}_{\xi}$ acts on $\partial V$ like on a $(1,1)$ 
tensor, while the second term is the anomalous term reflecting that $\partial V$ is in fact not a tensor.     
In index-free notation (\ref{ANOmalous}) reads 
 \be\label{anomalousdV}
  \delta_{\xi}\big(\partial V\big) \ = \ {\cal L}_{\xi}\big(\partial V\big) - \vec{V}\big(\partial \xi\big)\;. 
 \ee 
Below we will need the bilinear symmetric operation  
 that acting on two vectors gives a function: 
 \be
  \varphi(V,W) \ \equiv \ {\rm tr}(\partial V\cdot\partial W)
   \ \equiv \  \partial_i V^j \,\partial_j W^i \;.   
 \ee 
We stress that while $\varphi(V, W)$ has no free indices, it is \textit{not} a scalar built from $V$ and $W$.  
As the notation suggests, $\varphi(V, W)$ can be viewed as the trace of the matrix product of the
matrices $\partial V$ and $\partial W$.   
In terms of this symmetric function the deformed inner product (\ref{INNERMORE}) 
becomes
 \be\label{covINNER}
  \langle V+\tilde{V} | W+\tilde{W}\rangle'  \ = \ 
   i_V\tilde{W}+i_W\tilde{V}-\varphi(V,W)   \ = \  
  \langle V+\tilde{V} | W+\tilde{W}\rangle-\varphi(V,W) \;. 
 \ee
We also need an object $\tilde \varphi (V, W)$  that, given two vectors $V$ and $W$, gives a `one-form'  
  \be
  \tilde \varphi(V,W) \ \equiv \ {\rm tr}(\partial_i \partial V\partial W) \, dx^i 
   \ \equiv \  (\partial_i \partial_k V^j) \,\partial_j W^k \, dx^i \;,
      \ee 
where the $\partial_i$ does not interfere with the trace operation.   In components we
write
\be\label{tildephicomp}
\tilde \varphi_i(V,W) \ =\ {\rm tr}(\partial_i \partial V\partial W) \;. 
\ee
Note that while still bilinear,  $\tilde \varphi (V,W)$  is not symmetric under the exchange of $V$ and $W$, as the extra derivative
associated with the one form acts on the first vector.  
We can now write 
the deformed Lie derivative on a one-form in (\ref{corrliegen}) as 
 \be\label{DEFLie9}
 \  \delta'_{\xi+\tilde{\xi}}\,\tilde{V}   
 \ = \ {\bf L}_{\xi+\tilde{\xi}}\, \tilde{V} \ \equiv \ 
  {\cal L}_{\xi}\tilde{V}-i_V{\rm d}\tilde{\xi}\, - \,\tilde\varphi(\,\xi,V)
   \ = \ \widehat {\cal L}_{\xi+ \tilde \xi} \tilde V - \,\tilde\varphi(\,\xi,V)\;.    
 \ee 
For the vector and one form parts taken together we have   
 \be
\label {DEFLie}
\begin{split}
\phantom{\biggl(}  \delta'_{\xi+\tilde{\xi}}\,(V+ \tilde{V} )\ = \ & \ 
 {\bf L}_{\xi+\tilde{\xi}}\, (V +\tilde{V}) \ = \ 
  {\cal L}_{\xi} (V + \tilde{V})  \, -i_V{\rm d}\tilde{\xi}\, - \,\tilde\varphi(\, \xi,V) \\
 \ = \ & \ \widehat {\cal L}_{\xi + \tilde \xi} (V + \tilde V)
 - \,\tilde\varphi(\, \xi,V)  \;. 
 \end{split} 
 \ee 
Recognizing that the undeformed variations are
\be\label{DEFLiep}
  \delta_{\xi+\tilde{\xi}}\,(V+ \tilde{V} )\ =  \ 
  {\cal L}_{\xi} (V + \tilde{V})  \, -i_V{\rm d}\tilde{\xi}\, ,   
 \ee
we can write  
 \be
 \label{exp1} 
 \delta'_{\xi+\tilde{\xi}} \ = \ \delta_{\xi+\tilde{\xi}} \ +\  \tilde \delta_{\xi+\tilde{\xi}} \;,
\ee
 with
 \be
 \tilde\delta_{\xi+\tilde{\xi}}\,(V+ \tilde{V} )\ = - \,\tilde\varphi(\, \xi,V)\;.    
 \ee 
Finally, the deformed Courant bracket (\ref{DEfCbracket})
can also be written neatly using $\tilde \varphi$:
  \be
 \begin{split}
\phantom{\Bigl(}   \big[\, V+\tilde{V}, W+\tilde{W} \,\big]'  \ = \ \  &
  \big[\,V\,,W\, \big]\, +\, {\cal L}_{V}\tilde{W}-{\cal L}_{W}\tilde{V}
  -\f{1}{2} {\rm d} \big(i_{V}\tilde{W}-i_W\tilde{V}\big) \ \  \\
  &\hskip-8pt -\f{1}{2}\big(\tilde\varphi(V,W)-\tilde\varphi(W,V)\big)\;. 
 \end{split}  
 \ee 
Of course, we also have 
\be
 \big[\, V+\tilde{V}, W+\tilde{W} \,\big]'  \ = \   \big[\, V+\tilde{V}, W+\tilde{W} \,\big]
  -\f{1}{2}\big(\tilde\varphi(V,W)-\tilde\varphi(W,V)\big) \,,
\ee
where the first term on the right-hand side is the original Courant bracket.  The
$B$ automorphism also holds: 
 \be
    \big[\, V+\tilde{V}+i_VB\, , \, W+\tilde{W}+i_WB \,\big]' \ = \ \big[\, V+\tilde{V}\, , \,
    W+\tilde{W} \,\big]'\,
    +\, i_{[V,W]} B \;, 
 \ee 
as can be easily verified directly.  

Let us now prove that the deformed inner product (\ref{covINNER}) transforms covariantly 
under the deformed Lie derivative (\ref{DEFLie}), i.e., 
 \be
  \delta'_{\xi+\tilde{\xi}}\, \langle V+\tilde{V} | W+\tilde{W}\rangle' \ = 
  \ {\bf L}_{\xi+ \tilde \xi} \,   
  \langle V+\tilde{V} | W+\tilde{W}\rangle'  
  \ = \ {\cal L}_{\xi} \langle V+\tilde{V} | W+\tilde{W}\rangle'\;,
 \ee 
where the second equality holds because on scalars the deformed Lie derivatives
are defined to act as ordinary ones.  Using the expansion (\ref{exp1}) 
and noting that the original inner product
is covariant under the standard Lie derivatives,  we get the condition:
\be
  \delta_{\xi+\tilde{\xi}}\, (-\varphi (V, W) ) 
  + \tilde \delta_{\xi+\tilde{\xi}}\, \langle V+\tilde{V} | W+\tilde{W}\rangle'  
   \ = \  - {\cal L}_{\xi}
   \varphi (V, W) \;. 
 \ee 
Since $\tilde \delta$ does not act on vectors, we can delete the prime on
the second term of the left-hand side and get
\be
-\varphi ({\cal L}_\xi V, W)   -\varphi ( V, {\cal L}_\xi W)  
  + \, i_V (-\tilde\varphi ( \xi, W))    + i_W  (-\tilde
  \varphi (\xi, V ) )
   \ = \  - {\cal L}_{\xi}
   \varphi (V, W) \;. 
 \ee 
Reordering the terms we find that this requires 
 \be\label{nonTensor}
  {\cal L}_{\xi}\varphi(V,W)-\varphi({\cal L}_{\xi}V,W)-\varphi(V,{\cal L}_{\xi}W) \ = \ 
  i_V\tilde \varphi(\xi,W)+ i_W\tilde\varphi(\xi,V)\;.    
 \ee 
This equation encodes the fact that the pairing $\varphi$ is non-tensorial.   
However, by virtue of this relation, the full inner product (\ref{covINNER}) \textit{is} tensorial (in fact, a scalar) in the 
deformed sense. The proof of (\ref{nonTensor}) is straightforward. 
Writing
$\Delta_{\xi}\equiv \delta_{\xi}-{\cal L}_{\xi}$, we have by (\ref{anomalousdV}) that 
$\Delta_{\xi}(\partial V)=-\vec{V}(\partial \xi)$. We thus  
compute 
 \be
 \begin{split}
  \Delta_{\xi}\varphi(V,W) \ &= \ {\rm tr}\big(\Delta_{\xi}(\partial V)\,\partial W\big)
  +{\rm tr}\big(\partial V\Delta_{\xi}(\partial W)\big) \\
  \ &= \ -{\rm tr}\big(\vec{V}(\partial \xi)\,\partial W\big)-{\rm tr}\big(\partial V\vec{W}(\partial\xi)\big)\\
  \ &= \ -V^k{\rm tr}\big(\partial_k(\partial \xi)\,\partial W\big)
  -W^k{\rm tr}\big(\partial_k(\partial \xi)\,\partial V\big)\\
  \ &= \   -i_V\tilde \varphi(\xi,W)- i_W\tilde \varphi(\xi,V)\;. 
 \end{split}
 \ee 
Recognizing that the left-hand side of (\ref{nonTensor}) is by definition $-\Delta_{\xi}\varphi(V,W)$,  
this completes the proof of (\ref{nonTensor}) and thus of the covariance of the inner product (\ref{covINNER}) 
in the deformed sense.   

We now want to establish that
the deformed bracket $[ \cdot \, , \cdot ]'$  transforms covariantly in the deformed sense (\ref{DEFLie}), i.e., 
 \be\label{EStablish}
  \delta'_{\xi+\tilde{\xi}}\,  \big[\, V+\tilde{V}, W+\tilde{W} \,\big]' \ = \ \widehat{\cal L}_{\xi + \tilde \xi } \big[\, V+\tilde{V}, W+\tilde{W} \,\big]'
 \, -\, \tilde \varphi(\, \xi,[V,W])\;.
 \ee 
The covariance of the Courant bracket gives us
 \be\label{EStablish1}
  \delta_{\xi+\tilde{\xi}}\,  \big[\, V+\tilde{V}, W+\tilde{W} \,\big] \ = \ \widehat{\cal L}_{\xi + \tilde \xi } \big[\, V+\tilde{V}, W+\tilde{W} \,\big]
 \, \;, 
 \ee 
and therefore condition (\ref{EStablish})  requires  
\be\label{EStablish9}
\begin{split}
 -\f{1}{2} \delta_{\xi+\tilde{\xi}}\,   \big(\tilde\varphi(V,W)-\tilde\varphi(W,V)\big)
  +  \tilde\delta_{\xi+\tilde{\xi}}\,  \big[\, V+\tilde{V}, W+\tilde{W} \,\big]\ = \  & \ \\
-\f{1}{2} {\cal L}_{\xi}\,   \big(\tilde\varphi(V,W)
\,  -\tilde\varphi(W,V)\big) \,  & -\, \tilde \varphi(\, \xi,[V,W])\;.
\end{split}
 \ee 
Writing out the variations this becomes 
\be\label{EStablish99}
\begin{split}
& \  -\f{1}{2} \,   \big(\tilde\varphi({\cal L}_\xi V,W)
 + \tilde\varphi( V,{\cal L}_\x W)\big) - {\cal L}_{V}\varphi(\xi,W)+ \f{1}{2}{\rm d}\big(i_V\varphi(\xi,W)\big)
 - (V \leftrightarrow W)  \\
&\qquad \qquad  = \ -\f{1}{2} {\cal L}_{\xi}\,   \big(\tilde\varphi(V,W)
\,  -\tilde\varphi(W,V)\big) \,  -\, \tilde \varphi(\, \xi,[V,W])\;.
\end{split}
 \ee 
We now can reorganize it as follows: 
\be\label{EStablish99}
\tilde{\varphi}(\xi,[V,W]) \ = \  \f{1}{2} {\Delta}_{\xi}  \tilde\varphi(V,W)
+{\cal L}_{V}\tilde\varphi(\xi,W)- \f{1}{2}{\rm d}
\big( i_V\tilde \varphi(\xi,W)\big)   - (V \leftrightarrow W) \;.  \\
 \ee 
This relation can be proved by a direct computation, 
whose details we present in Appendix B.  This completes our proof 
of the covariance of the deformed Courant bracket.

\section{Discussion and Outlook} 

We have shown that the unusual gauge transformations of the $b$-field required in the 
Green-Schwarz mechanism 
find a  
geometric description in an extension of generalized
geometry.  This extension is defined by a fully consistent $\alpha'$ deformation
of the Courant bracket, found in the context of a C-bracket deformation
in double field theory~\cite{Hohm:2013jaa}. 
 It was explained there that this is the unique field-independent 
 $\alpha'$ deformation of the C-bracket.  It is
 likely that the associated field-independent deformation of the Courant bracket is also unique.  
 
 In the standard approach, the Green-Schwarz transformations of the $b$-field
 are unusual Lorentz rotations.  One must include $b$-field gauge transformations
 to close the Lorentz transformations.  As we have shown,   
 working with diffeomorphisms and $b$-field gauge transformations, the same physics
 results in a gauge algebra identified with a 
 field-independent deformation of the Courant bracket.
The realization of 
a deformed 
diffeomorphism symmetry on the $b$-field is novel. The 
deformed Courant bracket is  
covariant under suitable $\alpha'$ corrected 
diffeomorphisms.  
The Jacobiator of the deformed Courant bracket is an exact one-form,
and the $B$-shift automorphism of the original bracket is  preserved. 
These properties of the Courant bracket are guaranteed by the work 
in  \cite{Hohm:2013jaa} but were explained here with suitable notation
that does not use doubled coordinates.
The utility of the doubled formalism is that it allows one to construct 
gauge invariant actions with 
$\alpha'$ corrections
that are exactly T-duality invariant. 

It is known that natural classical formulations of string theory 
make use of elements of the theory that are usually understood as requirements
of the quantum theory.  For example, free string field theory, which is clearly
consistent in any dimension, uses a BRST operator that is only nilpotent
in the critical dimension.   
Similarly, 
in this note we showed that 
the modifications of the 
$b$-field gauge transformations, originally  
required by the cancellation of
a quantum anomaly, appears as part of the $\alpha'$ geometry of the
classical theory.  This is in accord with the discussion of \cite{Hull:1985jv,Sen:1985tq} that showed
that the unusual $b$-field transformations are needed to cancel a one-loop
anomaly of the {\em world-sheet} theory of heterotic strings.     

This work began as an investigation of the gauge transformations
of the theory described in  \cite{Hohm:2013jaa}   through a perturbative
identification of the metric and $b$-fields (section \ref{pert-clues}). 
It can be seen that the $\alpha'$ 
corrections of this theory violate the $b \to - b$ symmetry of bosonic string 
theory.  Thus~\cite{Hohm:2013jaa} does not describe a subsector of bosonic
closed strings, as originally expected, but rather a subsector of heterotic strings.
While heterotic strings are oriented string theories they do not have 
a $b \to - b$ symmetry.   The chiral CFT introduced in \cite{Siegel:1993th}
and further developed in \cite{Hohm:2013jaa} thus seems to have an anomaly 
that is not a feature of bosonic strings.   
As it does not include the familiar Riemann-squared
corrections but rather terms required by anomaly considerations,
 it appears to be a theory of `topological' type.

Given this result, 
how does one describe the $\alpha'$ corrections of bosonic
strings or heterotic strings \cite{Cai:1986sa,Gross:1986mw}, which include,
among others 
 the square of the Riemann tensor?  
Such an extension requires further 
deformations of the gauge structure of the two-derivative theory. 
In \cite{HZupcoming} we will report on  
a perturbative analysis of closed bosonic string field theory, which leads to the 
cubic action of ${\cal O}(\alpha')$.   To that order, the
gauge algebra is a  deformation of the C-bracket that involves
background values of the generalized metric.

The natural language needed to discuss
the action of 
deformed diffeomorphisms, especially `large' ones, is yet to be
developed.  
One  needs an extension of generalized geometry that 
 incorporates the $\alpha'$ deformed symmetry structures  for  
the action on one- and two-forms, possibly extending the theory of gerbes. 
In DFT a first step would be to find a 
finite form of the $\alpha'$ corrected 
generalized diffeomorphisms, 
extending those given in \cite{Hohm:2012gk} and 
studied in \cite{Hohm:2013bwa,Berman:2014jba,Park:2013mpa,Hull:2014mxa}.   
A more complete picture should arise upon inclusion of the Riemann squared
and other $\alpha'$ corrections into the structure.

\noindent
\textit{Note added:} 
At the completion of this work the paper \cite{Bedoya:2014pma}
appeared, which aims to describe first-order $\alpha'$ corrections of heterotic 
string theory in DFT.   In this construction 
the generalized Lie derivatives are not $\alpha'$-deformed, but the duality group
is extended.

\section*{Acknowledgments}
This is work is supported by the 
U.S. Department of Energy (DoE) under the cooperative 
research agreement DE-FG02-05ER41360. 
The work of O.H. is supported by a DFG Heisenberg fellowship. 
We thank Chris Hull and Warren Siegel 
for earlier collaborations
and subsequent discussions.  We happily acknowledge
several useful comments and suggestions by Ashoke Sen.    

\section*{Appendix}

\begin{appendix}

\baselineskip 15pt

\section{Comments on finite gauge transformations}
In this appendix we discuss some subtleties of the deformed diffeomorphisms that arise once we 
consider finite transformations.  
The Christoffel symbols transform under arbitrary general coordinate transformations
as 
 \be\label{transCrist}
  \Gamma_{mn}^{\prime \, k}(x')  \ = \ \frac{\partial x^p}{\partial x'^m}\frac{\partial x^q}{\partial x'^n}
  \frac{\partial x'^k}{\partial x^l}\Gamma_{pq}^{\,l}(x)+\frac{\partial x'^k}{\partial x^l}
  \frac{\partial^2 x^l}{\partial x'^m\partial x'^n}\;. 
 \ee 
It is convenient to introduce matrix notation, 
 \be 
 U^m{}_n \ = \ \frac{\partial x^m}{\partial x'^n}\;, 
  \qquad (U^{-1})^m{}_n \ = \ \frac{\partial x'^m}{\partial x^n}\;, 
 \ee 
so that the transformation (\ref{transCrist}) can be written as 
 \be\label{matrixtransGamma}
  {\Gamma}_m'(x') \ = \ U^n{}_{m} \left(\,U^{-1}{\Gamma}_n(x)U+U^{-1}\partial_n U\,\right)\;,
 \ee 
 and  in one-form notation,  $\Gamma' = \Gamma'_m dx'^m$ and $\Gamma = \Gamma_m dx^m$, 
 \be\label{matrixtransGamma-simpler}
  {\Gamma}' (x') \ = \ U^{-1}{\Gamma}(x)U+U^{-1}{\rm d}  U\, . 
   \ee 
Equation (\ref{CS-trans}) can be used to relate the CS forms of $\Gamma'$ and $\Gamma$:
\be
\label{CS-trans-vm}
\Omega(\Gamma') \ = \  \Omega (\Gamma)  \, - \, 
{\rm d}  \, \hbox{tr} \,\bigl(  {\rm d} U\, U^{-1} \wedge\Gamma  \bigr) - \f{1}{3} \, \hbox{tr} [ (U^{-1} {\rm d} U)^3 ]  \,.
\ee
The gauge invariance of $\widehat H$ requires that $b$ transforms in such a way that
\be
{\rm d} b'  + \f{1}{2} \Omega (\Gamma') \ = \ {\rm d} b + \f{1}{2} \Omega(\Gamma) \;, 
\ee
which gives
\be
{\rm d} b'  \ = \ {\rm d} \Bigl(  \,b + \f{1}{2}  \, \hbox{tr} \,\bigl(  {\rm d} U\, U^{-1}
\wedge \Gamma \bigr) \Bigr)
+ \f{1}{6} \hbox{tr} \big[ (U^{-1} {\rm d} U)^3 \big]\;. 
\ee
The last term is a closed three-form,  invisible for infinitesimal 
 transformations $x'^m=x^m-\xi^m(x)$, for which $U={\bf 1}+\partial\xi+{\cal O}(\xi^2)$.
Integrated over a three-manifold it yields the winding number of $U$.  Locally we 
write it as the exterior derivative of a two-form $j$: 
\be
w(U) \equiv -\f{1}{3} \, \hbox{tr} [ (U^{-1} {\rm d} U)^3 ] \ = \ {\rm d} j  \,, \qquad  j \ \equiv \ \f{1}{2} j_{ij}\, 
dx^i \wedge dx^j \,. 
\ee
With this we can conclude that the $b$ field transformation is given by
\be
b' \ = \ b \, + \, \f{1}{2}  \, \hbox{tr} \,\bigl(  {\rm d} U\, U^{-1}
\wedge \Gamma \bigr)\, - \, \f{1} {2} j \,. 
\ee
To linearized order in infinitesimal diffeomorphisms,  
for which we can ignore the last term, 
this indeed reduces to (\ref{nonlinDiff2v2}). 
In component notation the above equation gives
 \be\label{DEfDIff}
  b_{mn}'(x') \ = \ U^p{}_m U^q{}_{n}
  \Big(b_{pq}(x)+ {\rm tr}\big(\partial_{[p}U\,U^{-1}\,{\Gamma}_{q]}\big)
  - j_{pq}\Big)\;.  
 \ee 
Writing out the explicit derivatives 
yields 
 \be
  b_{mn}'(x') \ = \  \frac{\partial x^p}{\partial x'^m}\frac{\partial x^q}{\partial x'^n}\Big(b_{pq}(x)
  + \, \frac{\partial x'^r}{\partial x^k}\frac{\partial^2 x^l}{\partial x'^r\partial x'^s}
  \frac{\partial x'^s}{\partial x^{[p}}\,\Gamma_{\,q]l}^{\,k}-j_{pq}\Big)\;. 
 \ee 
This way of achieving 
gauge covariance under finite or large transformations is completely analogous to the Yang-Mills 
modification that is present already for the two-derivative $N=1$, $D=10$ supergravity \cite{Chapline:1982ww}. 
For general 
finite or large diffeomorphisms it would be useful to have
a  closed form expression for the two-form $j_{mn}$.

\section{Technical details for proof of covariance}
In this appendix we explicitly prove equation (\ref{EStablish99}), 
needed 
to establish the covariance of the deformed Courant bracket under deformed 
diffeomorphisms.  First we need a few relations. 
An explicit computation in local coordinates 
shows
 \be
  \Delta_{\xi}\big(\partial_i\partial_lV^k\big) \ = \ \partial_i\partial_l\xi^p \partial_pV^k
  -\partial_i\partial_p\xi^k \partial_lV^p-\partial_i\partial_l\partial_p\xi^k V^p
  -\partial_l\partial_p\xi^k \partial_iV^p\;. 
 \ee 
In matrix notation this can be written as 
 \be
  \Delta_{\xi}\big(\partial_i (\partial V)\big) \ = \ \partial_i (\partial\xi)\cdot \partial V
  -\partial V \partial_i (\partial\xi)-\vec{V}\big(\partial_i (\partial\xi)\big)
  -(\partial_i \vec{V})(\partial\xi)\;. 
 \ee  
We then compute for the first term on the r.h.s.~of (\ref{EStablish99}),  using (\ref{tildephicomp}),  
  \be\label{Result1}
 \begin{split}
  \Delta_{\xi}\tilde\varphi_i(V,W) \ &= \ \Delta_{\xi}{\rm tr}\big( \partial_i \partial V\, \partial W\big) 
  \ = \ {\rm tr}\big(\Delta_{\xi}(\partial_i \partial V) \partial W\big)+ {\rm tr}\big(\partial_i \partial V \Delta_{\xi}(\partial W) \big)\\
  \ &= \ {\rm tr}\big(\partial_i (\partial\xi)\partial V\partial W
  -\partial V \partial_i (\partial\xi)\partial W-\vec{V}\big(\partial_i (\partial\xi)\big)\partial W\\
   &\qquad\;\; -(\partial_i \vec{V})(\partial\xi)\partial W-\partial_i\partial V\vec{W}(\partial\xi) \big)
   \;. 
 \end{split} 
 \ee 
For the remaining terms on the r.h.s.~of (\ref{EStablish99}) we first note, 
recalling that ${\cal L}_V={\rm d}i_V+i_V{\rm d}$ on forms,  
   \be
   \label{B4-}
{\cal L}_{V}\tilde \varphi(\xi,W)  -    \f{1}{2} {\rm d} i_V\tilde \varphi(\xi,W) \ = \ 
  \f{1}{2} {\rm d} i_V\tilde \varphi(\xi,W)\, + \, i_V{\rm d}\tilde \varphi(\xi,W)\;. 
 \ee 
Next we compute 
the two terms on the right-hand side of this equation.
For the first one,  
 \be\label{Result2}
 \begin{split}
  \f{1}{2} {\rm d} i_V\tilde \varphi(\xi,W) \ &= \ \f{1}{2}  \partial_i\,{\rm tr}\big(\vec{V}(\partial\xi)\partial W\big)dx^i  \\
  \ &= \ \f{1}{2}{\rm tr}\big((\partial_i \vec{V})(\partial\xi)\partial W+\vec{V}(\partial_i(\partial\xi))\partial W
  +\vec{V}(\partial\xi)\partial_i(\partial W)\big)dx^i\; .
 \end{split}
 \ee
For the second one we have
 \be
 \begin{split}
  i_V{\rm d}\tilde\varphi(\xi,W)  \ &=  \   i_V{\rm d} \,  \bigl( {\rm tr} ( (\partial_j \partial\xi) \partial W) dx^j \bigr)  \ = \  i_V   [ \partial_i \bigl\{ {\rm tr} ( (\partial_j \partial\xi )\partial W) \bigr\}   dx^i \wedge dx^j  ]   \\
    \ &= \ i_V   [   {\rm tr} ( (\partial_j \partial\xi )\partial_i\partial W) 
  dx^i \wedge dx^j  ] \ =    {\rm tr} ( (\partial_j \partial\xi )\partial_i\partial W) 
   (V^i  dx^j - dx^i V^j ) 
  \\
   \  &= \  {\rm tr} ( (\partial_i \partial\xi )\vec{V}\partial W
  -(\vec{V} \partial\xi ) \partial_i\partial W) \, dx^i  \;.  \\
  \end{split}
  \ee
Back in (\ref{B4-})    
 \be\label{Result3}
 \begin{split}
  {\cal L}_{V}\tilde \varphi(\xi,W)  -    \f{1}{2} {\rm d} i_V\tilde \varphi(\xi,W) \ = \ 
  {\rm tr}\Big(&\f{1}{2}(\partial_i \vec{V})(\partial\xi)\partial W+\f{1}{2}\vec{V}(\partial_i(\partial\xi))\partial W\\
  &\;\;-\f{1}{2}\vec{V}(\partial\xi)\partial_i(\partial W)+\partial_i(\partial\xi)\vec{V}(\partial W)\Big)dx^i\;. 
 \end{split} 
 \ee 
Inserting now (\ref{Result1}) and (\ref{Result3}) into the right-hand side of (\ref{EStablish99})
we find after a quick computation 
 \be\label{FinalStep}
  {\rm r.h.s.}(\ref{EStablish99}) \ = \  {\rm tr}\big(\f{1}{2}
  ( \partial V\partial W -\partial W\partial V )  
  \partial_i(\partial\xi)
  +\partial_i (\partial\xi)\vec{V}(\partial W)\big)-(V\leftrightarrow W)\;, 
 \ee 
where we used repeatedly the antisymmetry in $(V\leftrightarrow W)$ and the cyclic property  of the trace. 
On the other hand, we compute for the left-hand side of (\ref{EStablish99}) 
 \be
 \begin{split}
  \tilde\varphi_i\big(\xi, [V,W]\big)  \ &= \  {\rm tr}\big(\partial_i (\partial\xi)\partial[V,W]\big) 
  \ = \ \partial_i\big(\partial_k\xi^l\big)\partial_l
  \big(V^p\partial_pW^k-(V\leftrightarrow W)\big)   \\
  \ &= \ {\rm tr}\big(\partial_i(\partial\xi)\partial V\partial W+\partial_i(\partial\xi)\vec{V}(\partial W)\big)-(V\leftrightarrow W)\;. 
 \end{split}
 \ee 
Due to the antisymmetrization in $(V\leftrightarrow W)$ this equals (\ref{FinalStep}). This completes the 
proof of (\ref{EStablish99}) and thus establishes 
the covariance of the deformed Courant bracket.

\end{appendix}

\end{document}